\documentclass{JHEP3}
\usepackage{epsfig}
\usepackage{graphicx}
\usepackage{dcolumn}
\usepackage{bm}
\newcommand{\be}{\begin{equation}}
\newcommand{\ee}{\end{equation}}
\def\bea{\begin{eqnarray}}
\def\eea{\end{eqnarray}}

 \def\be{\begin{equation}}
\def\ee{\end{equation}}
\def\bea{\begin{eqnarray}}
\def\eea{\end{eqnarray}}

\def\lesssim{\mathrel{\hbox{\rlap{\hbox{\lower4pt\hbox{$\sim$}}}\hbox{$<$}}}}
\def\gtrsim{\mathrel{\hbox{\rlap{\hbox{\lower4pt\hbox{$\sim$}}}\hbox{$>$}}}}

\title{P-term, D-term and F-term inflation}

\author{Renata Kallosh and Andrei Linde\\
    Department of Physics, Stanford University, Stanford, CA 94305,
USA\\
e-mail: kallosh@stanford.edu; alinde@stanford.edu}
\received{}       
 \preprint{SU-ITP-03/14\\  ~hep-th/0306058\\ June 6 2003}

\abstract{P-term inflation is a version of hybrid inflation which naturally appears in some brane inflation models. It was introduced in the framework of N=2 supersymmetric gauge theory where superconformal $SU(2,2|2)$ symmetry is broken down to N=2 supersymmetry by the vev of the auxiliary triplet field $\vec P$. Depending on the direction of this vev, one can get either D-term inflation or F-term inflation with a particular relation between Yukawa and gauge coupling, or a mix of these models. We show that F and D models, before  coupling to gravity is included, are related by a change of variables. Coupling of this model to N=1 supergravity breaks this symmetry and introduces a class of P-term models interpolating between D-term and F-term inflation. The difference between these models is determined by the direction of the vector $\vec P$, which depends on the fluxes in the underlying D3/D7 model of brane inflation. We discuss cosmological consequences of various versions of P-term inflation.}

\begin{document}

\section{Introduction}

Over the last few years there were many attempts to implement inflation in brane cosmology based on string theory, see for example a review in \cite{Quevedo:2002xw}. Evolution of the universe in these models is described by various versions of the hybrid inflation scenario \cite{Linde:1991km}. However, despite an impressive progress reached in this direction and a multitude of available models, several problems still remain unsolved. The most important ones are moduli stabilization, coupling to gravity, and the problem of exit from inflation.

{\it 1. Coupling to gravity.}~~~
Inflationary potentials appearing in all existing stringy brane inflation models were derived ignoring gravitational effects \cite{Quevedo:2002xw}. It was assumed that one should simply find the expression for the potential describing interacting branes and substitute it into the Einstein equations. However, it is well known that the moduli fields in supergravity often acquire masses $|m^2| = O(H^2) \sim V/M_p^2$, which make inflation very hard to achieve. These terms appear only after one takes into account gravitational effects. One can avoid such terms in D-term inflation \cite{Binetruy:1996xj,Halyo:1996pp} and in the  special  versions of F-term inflation \cite{Copeland:1994vg,Dvali:1994ms,Linde:1997sj}, but it is not always obvious whether the brane inflation models described  in \cite{Quevedo:2002xw} belong to any of these two classes.

{\it 2. Moduli stabilization.}~~~
The major assumption made in all of these models is that the dilaton and the volume of the compactified space are fixed by some unknown mechanism. One would like to find a version of brane inflation where the motion of branes responsible for the slow-roll inflation is consistent with the stabilization of the dilaton and the volume. This problem is so complicated that until very recently it was not known how to obtain de Sitter solutions in string theory. This problem was solved in \cite{Kachru:2003aw} where a de Sitter solution of type IIB string theory was obtained, with all moduli fields being stabilized. One would like to use this experience in stabilization of stringy moduli in application to inflation. We expect that the D3/D7 model of brane inflation developed in \cite{Herdeiro:2001zb,Dasgupta:2002ew} will be suitable for this purpose.

{\it 3. Exit from inflation.}~~~
In most of the models of stringy brane cosmology, the exit from the stage of inflation occurs due to tachyon condensation. The description of this effect is based on Sen conjecture \cite{Sen}. The process of tachyon condensation may lead to severe cosmological problems described in \cite{Kofman:2002rh}. The only known exception from this rule is the D3/D7 model of brane inflation developed in \cite{Herdeiro:2001zb,Dasgupta:2002ew}, where the exit from inflation and the subsequent process of reheating occurs in the same way as in the simplest models of D-term and F-term inflation \cite{Felder:2000hj}.

Out of these three problems, the problem of moduli stabilization  is the most complicated one; it will be considered separately \cite{inprogress}.  In this paper we will discuss the first of these problems (coupling to gravity) considering as an example the version of the hybrid inflation \cite{Linde:1991km} called  ``P-term inflation.''  This scenario was  introduced in  \cite{Kallosh:2001tm} in the context of N=2 SUSY. Later it was found that this scenario is realized in the D3/D7 model of Ref. \cite{Herdeiro:2001zb,Dasgupta:2002ew}. Interestingly, until one introduces gravity, this scenario simultaneously describes D-term {\it and } F-term inflation. 
The difference between these two realizations appears only when one breaks N=2 supersymmetry and implements P-term inflation in N=1 supergravity. 

The version of this scenario described in \cite{Herdeiro:2001zb,Dasgupta:2002ew} was based on D-term inflation. In this paper we will describe the F-term version of this theory. Moreover, as we will show, P-term inflation in the context of N=1 supergravity leads to a new class of inflationary models, which interpolate between D-term and F-term models. The choice of a particular model is determined by the choice of the vev of the auxiliary  triplet field $\vec P$, which, in its turn, is determined by the fluxes on the branes.

An additional motivation for considering the F-term implementation of the P-term model, as well as of the models interpolating between the D-term and F-term inflation, comes from the fact that the recent WMAP observations suggest a possibility of a small dependence  of the spectral index on momenta \cite{Peiris:2003ff}. It was pointed out in \cite{Peiris:2003ff}  that the model of Linde and Riotto \cite{Linde:1997sj} gives an example of such situation. This model is  an F-term inflation model \cite{Copeland:1994vg,Dvali:1994ms}  with a canonical K\"{a}hler potential.  It may be possible to justify the choice of a canonical K\"{a}hler potential in this model if one considers it as a brane inflation model in a compactified string theory \cite{inprogress}.  Note that F (D) are auxiliary fields of chiral (vector) multiplets of N=1 supersymmetry, whereas $\vec P$ is the triplet of auxiliary fields in N=2 vector multiplet.   

The N=1 generic D-term inflation \cite{Binetruy:1996xj,Halyo:1996pp} models have a special advantage over generic N=1 supersymmetric F-term inflation \cite{Copeland:1994vg,Dvali:1994ms}: the inflaton field in D-term models which have vanishing superpotentials and their first derivatives during inflation, $W=W'=0$, does not acquire a Hubble scale mass term from the supergravity corrections, i.e. the inflaton direction remains flat. On the other hand, the F-term model \cite{Linde:1997sj}, which may explain the running of the scalar spectrum index of the CMB fluctuations, also does not acquire such a mass term, whereas quartic and higher order terms are generated by supergravity corrections. Thus is would be interesting to clarify the relation between D-term and F-term inflation using the fact that they may be related to an N=2 supersymmetric model.

An attempt was already made to promote the  theory of F-term inflation  to N=2 supersymmetry in \cite{Watari:2000jh}. The main motivation of \cite{Watari:2000jh} was to remove the difficulty of the F-term dominated inflation related to large radiative corrections to the K\"{a}hler potential. However, the potential used in \cite{Watari:2000jh} does not have N=2 supersymmetry.  We will find here a correct version of the N=2 supersymmetric F-term inflation model; it will contain specific D-terms. 

We will start description of P-term inflation in Section 2 with a global $SU(2,2|2)$-superconformal gauge theory,  which corresponds to a dual gauge theory of supersymmetric D3/D7 branes.  The model has  one vector multiplet and one charged hypermultiplet. The non-vanishing auxiliary fields $\vec P $, a triplet of Fayet-Iliopoulos terms,   explicitly break the superconformal symmetry down to N=2 supersymmetry.

Using the symmetries of  the  P-term model we will show  that   the F-term  model, supplemented by proper D-terms, and the  D-term  model with $\lambda^2=  2 g^2 $ are equivalent, being related by a change of variables.

Then we will discuss what happens when one introduces coupling of these fields to  N=1 supergravity. It was explained in \cite{Kallosh:2001tm} that it is  not possible to embed the P-term model into N=2 supergravity. Coupling to N=1 supergravity makes a significant difference between F and D models, as we will see, despite the fact that in the limit when $M_{Pl}\rightarrow \infty$ the gauge models are equivalent. In particular, we will find that a generic $P$-term model coupled to N=1 supergravity is characterized  by an additional  parameter $0\leq f  \leq 1$. When $f =0$, the D-term model is recovered, for $f =1$ we find an F-term model. This dependence on $f $ disappears in the limit $M_{Pl}\rightarrow \infty$.

In Section 3 we  will study inflation in these models. We emphasize, following \cite{Kallosh:2001tm,Endo:2003fr}, that two different regimes are possible in P-term, D-term and F-term models. The regime often discussed in the literature is realized for relatively large values of coupling constants  \cite{Binetruy:1996xj,Halyo:1996pp,Copeland:1994vg,Dvali:1994ms,Linde:1997sj,Lyth:1998xn}. However, in this regime, in addition to the standard inflationary perturbations one also has large perturbations of metric  produced by cosmic strings \cite{Kofman:1987wm,Jeannerot:1997is}. To avoid the contradiction with the observational data one should suppress stringy contribution to the perturbation of metric.  One can do it, e.g., by adding new fields and new terms to the superpotential \cite{Jeannerot:2000sv,Lyth:1998xn}, or by considering alternative mechanisms of generation of density perturbations \cite{Endo:2003fr}. We show, however, that for a sufficiently small values of the coupling constant the standard results for density perturbations in P-term, D-term and F-term inflation should be reconsidered, which allows one to suppress string contribution without making any modifications of the original models. A distinguishing feature of this class of models in the weak coupling limit is exact flatness of the spectrum of density perturbations, $n = 1$.
 
\section{P-term model
}

\subsection{$SU(2,2|2)$ superconformal part of the model}

We have an abelian N=2 {\it   gauge multiplet} which belongs to a representations of rigid $SU(2)$, the antisymmetric tensor $ \varepsilon_{AB} $, $A,B=1,2$,  is used to raise and lower the $SU(2)$ indices. The multiplet consists of a complex scalar $\Phi_3$, a vector $A_\mu$, (all singlets in $SU(2)$),  a spin-1/2 doublet $ \lambda^{ A}= \varepsilon^{AB} \gamma_5 C \bar \lambda_B^T$ (gaugino) and an  auxiliary field
$P^{r}$, triplet in $SU(2)$.

A  hypermultiplet, N=2 {\it   matter multiplet} has two complex scalar fields forming a doublet under $SU(2)$, $\Phi^A$ and $\Phi_A= (\Phi^A)^*$   and a spin 1/2 field $\psi$, singlet under $SU(2)$ (hyperino). There is also a doublet of dimension 2 auxiliary fields, $F^A$ with $F_A= (F^A)^*$ .
The bosonic part of the superconformal action is
\bea
\label{model}
{\cal L}_{s.c.}&=&  D_\mu \Phi_3 D^\mu \Phi_3^*   -{1\over 4}F_{\mu\nu}^2 + {1\over 2} \vec P^2 \nonumber\\
\nonumber\\
&+&  D_\mu \Phi^A D^\mu \Phi_A  +  F^A F_A  \\
\nonumber \\
&+&  g   \Phi^A \vec \sigma _A{}^B \vec P \Phi_B - 2g^2 \Phi^A  \Phi_A \Phi_3 \Phi_3\ . \nonumber 
\eea
The covariant derivatives on the hypers are
\bea
D_\mu \Phi_A &=& \partial_\mu \Phi_A  +i g A_\mu \Phi_A \ , \nonumber\\
D_\mu \Phi^A&=& \partial_\mu \Phi^A -i g A_\mu \Phi^A \ .
\label{cov}\eea
The first line in (\ref{model}) is for the  vector multiplet, the second one is for the hypermultiplet. 
The third line  includes terms which must be added to the action simultaneously with covariantization of derivatives. All term  depending on $g$ describe the gauging. When the gauge coupling $g$ is vanishing, the theory is a non-interacting theory of a vector multiplet and a neutral hypermultiplet.

Note that all complex scalars have  kinetic terms of the form
$\partial \Phi \partial \Phi^*$  which simplifies the comparison with superfield notations\footnote{In \cite{Kallosh:2001tm} we used the notation with canonical kinetic terms of the form ${1\over 2}\partial \Phi \partial \Phi^*$.}.
Using equations of motion for auxiliary fields, 
\be
 \vec P=- g \Phi^A  (\vec \sigma)_A{}^B \Phi_B \ , \qquad  F^A=0 \ ,
\ee
we find the potential
\be
V_{s.c.}= 2 g^2\Bigl( \Phi^\dagger \Phi |\Phi_3|^2 +{1\over 4} (\Phi^\dagger \vec \sigma \Phi)^2\Bigr)\ ,\nonumber
\label{scpot}\ee
where  
$\Phi^\dagger \Phi\equiv \Phi^A \Phi_A=|\Phi_1|^2 + |\Phi_2|^2\geq 0$.

\subsection{P-term, F-term and D-term models}\label{equiv}
To break $SU(2,2|2)$ symmetry down to N=2 supersymmetry one can add the N=2 FI terms $\vec \xi$  
to the theory\footnote{To simplify the relation between F and D term  theories we changed notation: $\vec \xi$  here corresponds to    $-\vec \xi/ g$ in \cite{Kallosh:2001tm}. } (for abelian multiplet only). The N=2 supersymmetry of the action remains intact. The new action is:
\be
{\cal L}_{N=2}=
{\cal L}_{s.c.}- g \vec P  \vec \xi \ ,
\ee
where ${\cal L}_{s.c.}$ is presented in eq. (\ref{model}).
This will change the field equation for $P^r$, which will become
\be
\vec P= -g [(\Phi^\dagger \vec \sigma \Phi) - \vec \xi]\ .
\ee
The potential of the P-term model becomes equal to
\be
V_{N=2}^P= 2 g^2\left [\Phi^\dagger \Phi |\Phi_3|^2 +{1\over 4} \left(\Phi^\dagger \vec \sigma \Phi -\vec\xi  \right)^2\right]\ .
\label{compact}\ee
We may use the symmetries of the theory and point out the triplet $\xi^r$ in any direction. The potential can be given in the form suitable for N=1 notations. In the general case, we have 
\be
\xi \equiv \sqrt{|\vec \xi|^2}= \sqrt {\xi_+ \xi_- + (\xi_3)^2}\ , \qquad  \xi_{\pm} \equiv \xi_1\pm i\xi_2 \ ,
\ee
 and we will use  $S=\Phi_3$ for the neutral scalar, $\Phi_1=\Phi_+$ ($\Phi_2^*=\Phi_-$) for the positively (negatively) charged scalar   
\be
V_{N=2}^P= 2 g^2\left (|S \Phi_+|^2+ |S \Phi_-|^2  +\Bigl|\Phi_+ \Phi_- - {\xi_+ \over 2}\Bigr|^2 \right)+ {g^2\over 2}\Bigl(|\Phi_+|^2 -|\Phi_-|^2- \xi_3\Bigr)^2\ .
\label{general}\ee
Thus we find that P-term potential corresponds to an N=1 model 
\be
V= |\partial W|^2 + {g^2\over 2}D^2
\label{N1form}\ee
with a superpotential and a D-term
\be
W= \sqrt {2} g S(\Phi_+ \Phi_- - \xi_+/2) \ , \qquad D= |\Phi_+|^2 -|\Phi_-|^2- \xi_3 \ .
\ee
Note the potential of this N=2 supersymmetric gauge theory at $\xi_+=\xi_-=0$, $\xi_3=|\vec \xi|$ coincides with the particular case  of D-term inflation studied before in the cosmological context  in \cite{Binetruy:1996xj} with $W= \lambda S\Phi_+ \Phi_- $ and $ D= |\Phi_+|^2 -|\Phi_-|^2- \xi$
\be
V_{N=2}^D= 2 g^2\Bigl (|S \Phi_+ |^2+ |S \Phi_- |^2  +|\Phi_+ \Phi_-|^2 \Bigr)+ {g^2\over 2}\Bigl(|\Phi_+|^2 -|\Phi_-|^2- \xi\Bigr)^2\ .
\label{D}\ee
under condition that $\lambda = \sqrt{2}g$.
If we would choose $\xi_+=\xi_-=2M^2= \xi$ we would recover a potential of an F-term inflation model with $W= \lambda S(\Phi_+'\Phi_-'-M^2) $ and $  D= |\Phi_+'|^2 -|\Phi_-'|^2$
\be
V_{N=2}^F= 2 g^2\Bigl(|S \Phi_+' |^2+ |S \Phi_-' |^2  +|\Phi_+' \Phi_-' - M^2|^2 \Bigr)+ {g^2\over 2}\Bigl(|\Phi_+'|^2 -|\Phi_-'|^2\Bigr)^2\ .
\label{F}\ee 
The first term of the potential of our N=2 supersymmetric theory coincides  with the F-term inflation potential in N=1 theory, proposed in \cite{Dvali:1994ms}, under condition that  the gauge coupling $g$ defining $D$ terms in eq. (2) of \cite{Dvali:1994ms} and the Yukawa coupling $\lambda$ there are related, $\lambda= \sqrt{2} g$. The second term in (\ref{F}), a D-term, has to be added to the model in  \cite{Dvali:1994ms} to make it N=2 supersymmetric. This D-term vanishes during inflation, however, it is required by N=2 supersymmetry.
Note that the potential of N=2 theory, suggested to describe hybrid inflation and presented in eq. (1) of \cite{Watari:2000jh}, in the limit when $M_p\rightarrow \infty$
does not possess a global N=2 supersymmetry, since the last term in eq. (\ref{F}) is missing there. 

A nice property of the P-term model is that one can use $U(2)$ symmetry of the superconformal part of the action to perform a change of variables which related F-term version to D-term version and makes these two theories identical. To relate $V_{N=2}^F$ to $V_{N=2}^D$  we need to find an $U(2)$ transformation which will
accomplish that 
\bea
\Phi^\dagger  \sigma^3 \Phi &=& \Phi^{'\dagger}  \sigma^1 \Phi'\nonumber\\
\Phi^\dagger  \sigma^1 \Phi &=& \Phi^{'\dagger}  \sigma^3 \Phi'\nonumber\\
\Phi^\dagger  \sigma^2 \Phi &=& \Phi^{'\dagger}  \sigma^2 \Phi'\nonumber\\
\Phi^\dagger   \Phi &=& \Phi^{'\dagger}   \Phi'
\eea
Such transformation is indeed possible:
\be
\Phi_3= \Phi_3' \ , \qquad \Phi'_A = U_A{}^B \Phi_B \ , \quad  A=1,2 \ ,  \qquad U={1\over \sqrt 2} (\sigma^3+ \sigma^1)\ , 
\qquad U^\dagger U=I \ .
\ee
It can be verified, using eq. (\ref{compact}) that
\be
V_{N=2}^D(\Phi)= V_{N=2}^F(\Phi')
\ee
with 
\be
\Phi'_+= {1\over \sqrt 2}(\Phi_++ \Phi_-^*)\ , \qquad \Phi'_-= {1\over \sqrt 2}(\Phi_+- \Phi_-^*) \ .
\ee
Moreover,  the total action is $U(2)$ symmetric and therefore, when F-term model and D-term model are each promoted to N=2 supersymmetry, their gauge actions, not only the potential,  become identical up to a change of variables.


\subsection{Vacua of  P-term model}

For the vacuum solutions we chose the vanishing vector field and constant values of 3 complex scalar fields $\Phi_+, \Phi_-, S $. The potential (\ref{general}) has two minima.

1. {\it Non-supersymmetric local minimum with flat directions (de Sitter valley)} at some  undefined but restricted constant value of the scalar $S$, all other scalars vanish.
\be
\Phi_+=  \Phi_-=0   \ , \qquad |\vec P|^2= g^2\xi^2  \ , \qquad V_0= {1\over 2} g^2 \xi^2 \ .
\ee
This is a local minimum of the potential for $ |S|^2 > S_c^2\equiv \xi /2  $.  The curvature of the potential is positive when this restriction is applied. 
One finds that the all supersymmetries are broken  at this vacuum  as long as $|\xi|\neq 0$.  When coupled to gravity, this vacuum corresponds to  a {\it de Sitter solution}.
The mass spectrum of the hypers is split, the eigenvalues are
\be
M_\pm^2=  2 g^2 (  | S|^2 \pm   \xi/2) \  .
\ee
$S$ provides a  flat direction in the potential.

2. {\it Supersymmetric global minimum}
\be
S_{ \rm susy}=0 \ ,  \qquad  \vec P_{\rm susy}=-g(\Phi^\dagger \vec \sigma \Phi-\vec \xi) _{\rm susy} =0 \ , \qquad V_{\rm susy}= 0\ .
\ee
In components 
\be
|\Phi_+|^2- |\Phi_-|^2-\xi_3=0 \qquad |\Phi_+|^2+ |\Phi_-|^2=\xi \ .
\ee
It has a solution 
\be |\Phi_-|^2= { \xi- \xi_3 \over 2} \qquad |\Phi_+|^2= { \xi+ \xi_3 \over 2}\ee
In particular cases of D-term model with $\xi_+=0$ we recover $|\Phi_+|^2= \xi_3 $ and $\Phi_-= 0$ in the ground state. In F-term case with $\xi_3=0$ we find $|\Phi_-|^2=|\Phi_+|^2= { \xi\over 2}$.
The absolute minimum ground state has both N=2 supersymmetries unbroken and vanishing potential. The  gauge symmetry is spontaneously broken. There is one massive vector multiplet,  all fields have a mass $m^2= 2 g^2\xi$.

\subsection{Gauge theory loop corrections}

The flat direction of the inflaton field $S$ is 
uplifted    due to the first
loop corrections in gauge theory.  The tree level  splitting of the
masses in supermultiplets in de Sitter vacuum leads to the effective
1-loop potential for large inflaton field $S$ \, \cite{Dvali:1994ms,Kallosh:2001tm,Herdeiro:2001zb,Dasgupta:2002ew}:
\be
V_{1-\rm loop}= {g^2\xi^2\over 2}\left(1 + {g^2\over 8\pi^2} \ln {
|S^2|\over {|S_c^2|} } 
\right) .
\ee
This term is important because it leads to the motion of the field
$S$ towards the bifurcation point and the end of inflation. It is
interesting to note that for the N= 2 P-term model {\it
all non-gravitational higher loop corrections
are finite} \cite{Kallosh:2001tm}.


\subsection{Coupling of P-term model to N=1 supergravity}

Coupling of a  model with rigid N=2 supersymmetry to N=1 supergravity can be realized as follows.  We may ignore the fact that the rigid limit has double supersymmetry and proceed as if only one supersymmetry is available and make it local. This procedure is of course not unique.  We choose here the minimal K\"{a}hler potential for all 3 chiral superfields.
In the K\"{a}hler geometry all 3 chiral multiplets now enter in a symmetric way. 
Also the kinetic term function $f$ of the vector multiplet may be taken minimal, i.e. field independent delta-function. Thus our N=1 supergravity has the 
K\"{a}hler potential and the superpotential  given by
\be 
K= {|S|^2+|\Phi_+|^2 +|\Phi_-|^2\over M_p^2} \ , \qquad W= \sqrt {2} g S(\Phi_+ \Phi_- - \xi_+/2) \ .
\ee 
The complete supergravity potential takes a standard form defined by $K$ and $W$. We are interested in the effect of supergravity corrections at the period of inflation, when the charged fields have a vanishing vev and the inflaton field is large. In this regime  the potential is given by  the gauge potential (\ref{general}) and terms depending on  ${|S|^2\over M_p^2}$. It is useful here to note that for our superpotential
\be
W=S {\partial W\over \partial S}\ , \qquad K_S W = {S\bar S \over M_p^2}{\partial W\over \partial S} \ .
\ee
In the regime ${|\Phi_\pm |^2\ll M_p^2}$, the effective potential is given by 
\bea
V &=&  2g^2 e^{|S|^2\over M_p^2}  \left[
\Bigl|{\partial W\over \partial S}\Bigr|^2 \left (\Bigl(1 + {S\bar S \over M_p^2}\Bigr)^2 -3{S\bar S \over M_p^2}\right) +|S \Phi_+|^2+ |S \Phi_-|^2 \right] \nonumber\\
&+& {g^2\over 2}\Bigl(|\Phi_+|^2 -|\Phi_-|^2- \xi_3 \Bigr)^2  \ .
\eea
Note that the negative contribution due the term with the square of gravitino mass,  $ -3 e^{K\over M_P^2}\ {|W|^2 \over M_p^2} $ had an important role in this calculation as it is responsible for the negative  sign in front of the term quadratic in $S$.
The calculation of the supergravity corrections for the F-term part reproduce the result in \cite{Copeland:1994vg}, that the terms quadratic in inflaton mass cancel. This cancellation is a consequence of our choice of the minimal K\"{a}hler potential.  This fact is very important for the possibility to have realistic models of F-term inflation \cite{Linde:1997sj}, but it plays no role in D-term inflation because D-term is not affected by  the exponential SUGRA corrections due to  the K\"{a}hler potential.\footnote{It has been recently realized that there are some particular corrections for the D-term potential in supergravity, which disappear in the limit $M_p \to \infty$. These corrections, to be discussed in \cite{NewDterm}, do not affect the main conclusions of our work.}

We find for the complete potential with tree level supergravity corrections 
\bea
V &=&  2g^2 e^{|S|^2\over M_p^2}  \left[
|\Phi_+ \Phi_- - \xi_+/2|^2 \left (1 - {S\bar S \over M_p^2} +\Bigr({S\bar S \over M_p^2}\Bigl)^2\right) +|S \Phi_+|^2+ |S \Phi_-|^2 \right] \nonumber\\
&+& {g^2\over 2}\Bigl(|\Phi_+|^2 -|\Phi_-|^2- \xi_3 \Bigr)^2  \ .
\eea
The inflating trajectory takes place at $\Phi_+=\Phi_-=0$. If we add  
 1-loop gauge theory corrections, we find the following potential 
\be
V =  {g^2 \xi^2\over 2}\left(1 + {g^2\over 8\pi^2} \ln {
|S^2|\over {|S_c^2|} } +f   {|S|^4 \over 2 M_p^4} +\dots  \right), 
\ee
where 
\be
f  = {\xi_1^2+ \xi_2^2\over \xi^2}\ , \qquad 0 \leq f  \leq 1 \ ,
\ee
and $\dots$ stands for terms ${|S|^6 \over 2 M_p^6}$ and higher order gravitational corrections.  Special case $f =0$ corresponds to D-term inflation,\,  $f =1$  corresponds to F-term inflation. A general P-term inflation model has an arbitrary $0 \leq f  \leq 1$.

The running of the spectral index from blue to red was studied in \cite{Linde:1997sj}, it corresponds to having first the $|S|^4$ term driving inflation, and later, with smaller $S$, the log term driving inflation. In P-term model in general the contribution from supergravity has an additional parameter $f $, which allows to have a controllable running of the spectral index: no running D-term version or maximal running, F-term version, or any intermediate case with $0< f  < 1$.

It may be useful to mention here that in D3/D7 brane construction the three FI terms   $\vec \xi$  are provided by  a magnetic flux  triplet  $ \vec \sigma \, (1+\Gamma_5)\, F_{ab}\Gamma ^{ab} $, where $F_{ab}$ is the field strength of the vector field living on D7 brane in the Euclidean part of the internal space with $a=6,7,8,9$. The spectrum of D3-D7 strings depends only on $|\vec \xi|$ since there is no preferred direction for $\vec \xi$ in this construction.

\section{P-term inflation}

We will present here some  short remarks about the applications in
cosmology of N=2 supersymmetry  with P-term inflation
where $P^r$ is the triplet of the Killing prepotentials of N=2 gauge
theory. In this section it will be more convenient to switch from the fields $\Phi_i$ and $S$ to the  canonically quantized fields $\phi_\pm =\sqrt 2 \Phi_\pm$ and $s= \sqrt 2 S$. In terms of these fields the amplitude of the fields $\phi_\pm$ in the minimum of the potential, and the bifurcation point $s_c=\sqrt 2 S_c$, are given by
\be |\phi_-|^2= { \xi-\xi_3}, \qquad |\phi_+|^2=  \xi+ \xi_3, \qquad s_c^2 =\xi\ .
\ee
In particular cases of D-term model with $\xi_+=0$ we have $|\phi_-|^2= 2\xi $, $s_c^2 =\xi$, and $\phi_+= 0$ in the ground state. In F-term case with $\xi_3=0$ we find $|\phi_-|^2=|\phi_+|^2= s_c^2 ={ \xi}$.
The  effective potential, in units $M_p = 1$, is given by
\be\label{general2}
V =  {g^2\xi^2\over 2}\Bigl(1 + {g^2\over 8\pi^2} \ln {
|s^2|\over {|s_c^2|} } +f   {|s|^4 \over 8} +\dots \Bigr)\ .
\ee

\subsection{D-term case, $f  = 0$ }

\subsubsection{Inflation and adiabatic perturbations of metric}

To study inflation in this theory one should use the Friedmann equation
\begin{equation}\label{infl}
H^2 = \left({\dot a \over a}\right)^2 = V/3 \approx {g^2\xi^2\over 6},
\end{equation}
where $a(t)$ is a scale factor of the universe. Thus one has $H =
g \xi/\sqrt 6$.
This leads to inflation
\begin{equation}\label{infl1}
a(t) = a(0)~ \exp{ {{g \xi\  t \over\sqrt 6} }}~ .
\end{equation}
During the slow-roll regime the field $s = \sqrt{2} S$ obeys equation
$3H\dot s = -V'(s)$ \cite{book},
which gives
\begin{equation}\label{infl2}
s^2(t) = s^2(0) - {g^3\xi ~t\over 2\sqrt 6 \pi^2} \ .
\end{equation}

Assuming that $\xi < 1$, one can show that symmetry breaking occurs within the time much smaller than $H^{-1}$ after the field $s$ reaches the bifurcation point
$s_c =  \sqrt \xi$. Inflation takes place before the symmetry breaking. This is a generic property of
almost all versions of hybrid inflation \cite{Linde:1991km}. Using
equations (\ref{infl1}), (\ref{infl2}) one can find the value of the
field $s_{{}_N}$ such that the universe inflates $e^N$ times when the
field rolls from $s_{{}_N}$ until it reaches the bifurcation point $s = s_c$:
\begin{equation}\label{infl3}
s_{{}_N}^2  = {s_c}^2  + {g^2 N\over 2 \pi^2} = {\xi}  + {g^2 N\over 2 \pi^2}.
\end{equation}
Density perturbations on the scale of the  present cosmological horizon
have been produced at $s \sim s_{N}$ with $N \sim 60$, and their
amplitude is proportional to ${V^{3/2}\over V'  }$ at that time
\cite{book}. One can find parameters of our model using COBE
normalization for inflationary perturbations of metric on the horizon
scale \cite{Lyth:1998xn}:
\begin{equation}\label{smallg}
\delta_H ={1\over 5\sqrt 3 \pi} {V^{3/2}\over V'  }\approx 1.9\times
10^{-5} \ .
\end{equation}
In our model this yields
\begin{equation}\label{smallga}
 {V^{3/2}\over V'  }=  {2\sqrt 2\pi^2 \xi\over  g}~ s_{{}_N} \sim
5.2 \times 10^{-4},
\end{equation}
where $N \sim 60$.

From this point on, two different regimes are possible, depending on  the
value of the coupling constant $g$. The standard assumption made in the
literature is that $g^2$ is relatively large, ${g^2 N\over 2 \pi^2} \gg
{\xi}$ \cite{Lyth:1998xn}. In this case one has $s_{{}_N}   =  {g
\sqrt{N}\over \sqrt 2 \pi },$ and
\begin{equation}\label{COBE1}
 {V^{3/2}\over V'  }= 2\pi {\xi \sqrt N} \sim
5.2 \times 10^{-4}.
\end{equation}
For $N \sim 60$ this implies that
\begin{equation}\label{COBE2}
\xi  \approx  10^{-5}  \ .
\end{equation}
One can represent this result in terms of the amplitude of spontaneous
symmetry breaking of gauge symmetry. For the canonically normalized
fields  one has
\begin{equation}\label{COBE3}
|\phi_-| = \sqrt 2 s_c = \sqrt {2\xi} \approx 4.5 \times
10^{-3} M_p \approx \  10^{16}~ {\rm GeV}.
\end{equation}
This is very similar to the GUT scale. The spectrum of perturbations in
this model is nearly flat. It is characterized  by the spectral index
\begin{equation}\label{COBE4}
n = 1-{1\over N} \sim 0.98,
\end{equation}
which is in a  good agreement with the recent cosmological results \cite{Peiris:2003ff}.

The  results described above have been obtained for ${g^2 N\over 2 \pi^2} \gg {\xi}$, i.e. for  $ g
\gtrsim 2 \times 10^{-3}$. This is a natural assumption, but one should note
that $g^2$ in our model is not necessarily related to the gauge coupling constant in
GUTs, so it can be very small.
In the regime $g \ll 2\times  10^{-3}$ one has $s_{{}_N} \approx s_c=
\sqrt{\xi}$, so that
\begin{equation}\label{smallg2}
 {V^{3/2}\over V'  }= {2\sqrt 2\pi^2\over  g}\xi
^{3/2} \sim
5.2 \times 10^{-4}.
\end{equation}
This yields \cite{Kallosh:2001tm}
\begin{equation}\label{smallg3}
{\xi } \sim 7\times 10^{-4}~ g^{2/3}.
\end{equation}
Note that in this regime the expression for the spectral  index $n$
changes.  For $ g \ll 2 \times 10^{-3}$ one has exactly flat spectrum of
density perturbations,
\begin{equation}\label{COBE4a}
n = 1 \
\end{equation}
which is a distinguishing feature of this class of models.

\subsubsection{Cosmic strings}
When the field $s$ approaches the bifurcation point $s_c$, the  complex
field $\phi_- = |\phi_-| e^{i\theta} $ may roll in any direction $\theta$
in the process of spontaneous symmetry breaking. This process creates
cosmic strings \cite{Kofman:1987wm,Linde:1991km,Jeannerot:1997is}, which
have energy density $\mu = 2\pi\xi$ per unit length in our model,
corresponding to D-term model with $\lambda = \sqrt 2  g$ \cite{VilBook}.

Cosmic strings  lead to perturbations of metric proportional to $G\mu$,
in addition to the usual inflationary perturbations discussed  above.
The amplitude of these perturbations is of the same order of magnitude as
inflationary perturbations (\ref{COBE1}) in the usually considered regime
$ g \gtrsim 2 \times 10^{-3}$
\cite{Jeannerot:1997is,Linde:1997sj,Lyth:1998xn}. Ten years ago, this
would not be much of a concern, but recent observational data suggest
that the cosmic string contribution to density perturbations should be at
most at the level of one percent  of the inflationary contribution.
According to  \cite{Avelino:2003nn}, this implies that $G\mu  \lesssim
10^{-7}$, where $\mu = 2\pi\xi$ is the linear mass density of the
string.  This leads to a constraint
\begin{equation}\label{smallga2}
{\xi } \lesssim  4\times 10^{-7} \ .
\end{equation}
Restoring the Planck mass in our equations, one finds an upper  bound on
the amplitude of spontaneous symmetry breaking $|\phi_-|$ and on $s_c$
in D-term inflation (for its P-term related version with $\lambda = \sqrt
2 g$):
\begin{equation}\label{smallgb}
 |\phi_-| = \sqrt 2 s_c= \sqrt{2\xi}\lesssim 10^{-3} M_p \sim 2.4 \times 10^{15} {\rm GeV}  \ .
\end{equation}
Some authors argue that  $G\mu$  can be few times larger than  $10^{-7}$
\cite{Landriau:2003xf}, which would slightly relax the constraint on
$\xi$. However, even if the constraint on $\xi$ somewhat differs from
${\xi } \lesssim  4\times 10^{-7}$, it is difficult to make it consistent
with the requirement $\xi  \approx  10^{-5}$ (\ref{COBE2}) obtained for $
g \gtrsim 2 \times 10^{-3}$.

A possible resolution of this problem was proposed in
\cite{Kallosh:2001tm}.  Indeed, according to Eq. (\ref{smallg3}), the
required value of $\xi$ can be much smaller for  $ g \ll 2 \times
10^{-3}$ \cite{Kallosh:2001tm}. The condition ${\xi } \lesssim  4\times
10^{-7}$ implies that
\begin{equation}\label{smallgc}
g \lesssim 1.3 \times 10^{-5} \ .
\end{equation}
A similar constraint was obtained recently in  \cite{Endo:2003fr}.

In this simple model the requirement of suppression of string
contribution to perturbations of metric leads to the requirement that $g$
should be very small and, consequently, to a prediction of exact flatness
of the spectrum, $n = 1$, see Eq. (\ref{COBE4a}).

In this respect one should note that one can suppress the  contribution
of stings even if $g$ is large  by making some modifications of the
theory. For example, some modifications of the superpotential of F-term
inflation  lead to spontaneous breaking of $U(1)$ gauge symmetry already
at the stage of inflation. Therefore the cosmic strings  are diluted by
inflation \cite{Jeannerot:2000sv}. In D-term inflation the same goal can
be achieved by adding  more fields to the theory \cite{Lyth:1998xn}. It
is plausible that an analogous mechanism can be found in the general case
of  P-term inflation.

Another possibility is to take $\xi  \ll  10^{-5}$ to suppress
inflationary  and stringy perturbations (for arbitrary $g$), and then
consider an alternative mechanism of generation of density perturbations,
either using  \cite{Endo:2003fr} the curvaton mechanism
\cite{Linde:1996gt,Lyth:2001nq}, or using inflationary perturbations of
light scalars  to modulate the decay rate during preheating
\cite{Dvali:2003em}. In both cases, however, one would need to extend the
model by adding to it some new light scalar fields.

\subsection{F-term inflation, and a general case $f  \neq 0$}

Another interesting case to consider is F-term inflation, with  the
potential \be V =  {g^2\xi^2\over 2}\Bigl(1 + {g^2\over 8\pi^2} \ln {
|s^2|\over {|s_c^2|} } +{|s|^4 \over 8 } +\dots \Bigr) \ee This potential
differs from the D-term potential by the addition  of the term $\sim
s^4$. This term does not change the amplitude of stringy perturbations,
it practically does not change the value of $V$ during inflation, but it
increases $V'$, which makes inflationary perturbations somewhat smaller.

The change of $V'$ is small and the results coincide with the results
obtained for the D-term case if the first derivative of the term ${|s|^4
\over 8 }$ is much smaller than that of the term ${g^2\over 8\pi^2} \ln {
|s^2|\over {|s_c^2|} }$. This happens for $s \lesssim \sqrt g/3$. This
might seem to indicate that SUGRA corrections are important for small
$g$, when the one-loop effects are small. However, the situation is quite
different, because we are interested in the behavior of the potential at
$s \sim s_{{}_N}$, which is given by ${g\sqrt N\over \sqrt 2 \pi}$ for $g
\gtrsim 2 \times 10^{-3}$. Therefore $s_{{}_N} \gtrsim \sqrt g/3$ and
SUGRA corrections are important for large $g$. One can easily understand
it because at $g = O(1)$ one has $s_{{}_N} = O(1)$, in which case the
full SUGRA potential in F-term inflation blows up exponentially, and
inflation becomes impossible.

According to  \cite{Linde:1997sj}, for $g > 0.15$, inflation  in this
model is too short, whereas for $g \lesssim 0.06$ one can ignore the
SUGRA term $\sim s^4$ for the description of the last 60 e-folds of
inflation.  In the intermediate regime $0.06 < g < 0.15$ the spectrum of
perturbations in F-term inflation has some interesting features. It has a
maximum at an intermediate scale, and the spectral index $n$ runs from $n
< 1$ at small wavelengths to $n > 1$ at large wavelengths
\cite{Linde:1997sj}. Meanwhile, for $g \lesssim 0.06$ the last stages of
F-term inflation occur exactly as in the  D-term model.

However, this result is correct only if inflation occurs at  $s_{{}_N}
\gg s_c = \sqrt\xi$. Therefore for $g \lesssim 2 \times 10^{-3}$ these
results should be reconsidered in the same way as we did for D-term
inflation. In this regime the condition $s \ll \sqrt g/3$ implies that
$s_c^2 = \xi = 7\times 10^{-4}~ g^{2/3} \ll 10^{-1} g$, i.e $g \gg 3
\times 10^{-7}$. Thus, one can ignore the SUGRA corrections for $3 \times
10^{-7} \ll g \lesssim 0.06$, so all results obtained for D-term
inflation in the previous subsection remain valid for the F-term
inflation for $3 \times 10^{-7} \ll g \lesssim 0.06$. This means, in
particular, that one can ignore cosmic string effects for $3 \times
10^{-7} \ll g \lesssim 1.3 \times 10^{-5}$.

For $g \ll 3 \times 10^{-7}$ SUGRA correction dominate, the last 60
e-foldings occur at $s \approx  s_c = \sqrt\xi$, and therefore one has
\begin{equation}\label{smallgab}
 {V^{3/2}\over V'  }=  {g\sqrt 2 \over\sqrt \xi}\sim
5.2 \times 10^{-4},
\end{equation}
In this case ${\xi } \lesssim  4\times 10^{-7}$ and the cosmic string
effects are small for $g \lesssim 2\times 10^{-7}$. Thus for F-term
inflation, just as for D-term inflation, one can ignore contribution of
strings for $g \lesssim 1.3 \times 10^{-5}$, with a possible exception of
a small region at  $g \sim 3 \times 10^{-7}$.

In a more general case, interpolating between the D-term and  F-term
models, the SUGRA term $\sim s^4$ is multiplied by a small factor $f  <
1$ in the effective potential (\ref{general2}). Consequently, for small
$f$ the SUGRA effects are unimportant in a broader range of coupling
constants $g$, and the cosmic string contribution can be ignored for $g
\lesssim 1.3 \times 10^{-5}$.

As we mentioned above, F-term inflation models of the type of
\cite{Linde:1997sj}  lead to a running spectral index $n$ for $0.06 < g <
0.15$. On the other hand, in F-term inflation with $g \ll 2\times 10^{-3}$
the last 60 e-folds of inflation occur in a small vicinity of $s_c$, the
spectrum is almost exactly flat, $n = 1$.  Therefore if one wants to
obtain the running spectral index $n$ in the simplest F-term scenario,
one should use an alternative mechanism suppressing the contribution of
strings, see the previous subsection.

\section{Post-inflationary evolution and tachyonic preheating in P-term inflation}

The process of spontaneous symmetry breaking begins when the field $s$
reaches the bifurcation point $s_c = \sqrt \xi$. The further evolution of
the fields occurs practically independently of the small one-loop effects
and SUGRA corrections, so this process looks the same way (up to the
field redefinition described in Section \ref{equiv}) for all
implementations of P-term inflation, including F-term and D-term versions.

In particular, in the F-term scenario, the field combination  $\varphi =
\phi_+ -\phi_-$ remains equal to zero,  the absolute value of the
combination $\phi = \phi_+ +\phi_-$ grows from $0$ to $\sqrt {2 \xi}$,
whereas the field $s$ moves from $s_c$ to $0$. For definiteness, we will
assume that $\phi = |\phi| e^{i\theta}$ is real (see a discussion of a
more general case in the previous section).

For a generic hybrid inflation model, or for a general D-term model  with
two coupling constants $\lambda $ and $g$, the field trajectory after
spontaneous symmetry breaking is extremely complicated
\cite{Garcia-Bellido:1997wm}. An interesting and  somewhat surprising
property of the behavior of the fields during spontaneous symmetry
breaking in the F-term inflation is that, despite a very complicated
shape of the potential near the bifurcation point, the fields $\phi$ and
$s$ move along a straight line \cite{Bastero-Gil:1999fz}, \be {\phi\over
\sqrt 2} +s = \sqrt \xi \ . \ee Because of the equivalence of all
implementations of P-term inflation, all versions of P-term inflation,
including D-term inflation with $\lambda = \sqrt 2 g$, have this
interesting property.

Another unusual property of the process of oscillations of the fields
$\phi$ and $s$, which appears in all versions of hybrid inflation,
including P-term inflation, is that  the amplitude of coherent
homogeneous oscillations of the fields becomes extremely small {\it after
a single oscillation}, and the field distribution becomes completely
inhomogeneous. It consists of colliding waves of the classical fields
$\delta\phi \ll \phi$ and $\delta s \ll s$ on top of the spontaneous
symmetry breaking state with  $\phi = \sqrt {2\xi}$, $s = 0$.  This
happens because of the extremely rapid growth of quantum fluctuations
due to tachyonic instability (tachyonic preheating)  \cite{Felder:2000hj}.

\section{Conclusion}

In conclusion, we have presented above the P-term (N=2 supersymmetric)
gauge model  which unifies the F-term  and the D-term N=1 supersymmetric
models, and any combination of these two models, under the condition that
the Yukawa and the  gauge couplings are related as $\lambda=\sqrt {2} g$.
Coupling to N=1 supergravity differentiates between these models: they
are characterized by a free parameter $f $.  In a generic P-term
inflation scenario one has  $0\leq f  \leq 1$. The special cases $f = 0$
and $f = 1$ correspond to D-term inflation and  F-term inflation.

P-term inflation naturally appears in the context of the D3/D7 model of
brane inflation developed in \cite{Herdeiro:2001zb,Dasgupta:2002ew}.
Until the coupling to supergravity is introduced, one could expect that
the knowledge of the N=2 supersymmetric potential of the fields present
in the brane inflation model is sufficient for the description of
inflation. Our results indicate that these expectations may be correct
for the D-term version of P-term inflation, but in a more general case
$0< f  \leq 1$ the description of inflationary regime and its properties
may depend on the choice of the K\"{a}hler potential and on the value of the
parameter $f$, which is determined by the choice of fluxes on the branes.

As we pointed out in the beginning of the paper, in order to make the
brane  inflation models realistic, one should solve the problem of moduli
stabilization in the stringy brane inflation scenario. We hope that one
can resolve this problem by embedding these models into a more general
class of theories with stabilization of the dilaton and volume,
generalizing  an example of such stabilization in type IIB string theory
in \cite{Kachru:2003aw} for the case of inflation.

We are grateful to K. Dasgupta, S. Kachru, L. Kofman, J. Maldacena, S.
Prokushkin, S. Shenker, L. Susskind and S. Trivedi for stimulating
discussions.  We are especially thankful to T.~Moroi for important
comments. This work is supported by NSF grant PHY-9870115. The work by
A.L. was also supported by the Templeton Foundation grant No. 938-COS273.


\end{document}